\documentclass[%
 reprint,
superscriptaddress,
amsmath,amssymb,
aps,
pre,
]{revtex4-1}

\usepackage{graphicx}   
\usepackage{dcolumn}    
\usepackage{bm}  
\usepackage{xcolor} 

\usepackage{amsmath}
\usepackage{amssymb}
\usepackage{amsfonts,comment}
\usepackage{physics} 

\usepackage{mathtools}

\usepackage[colorlinks=true, linkcolor=blue, citecolor=red, urlcolor=red]{hyperref}

\newcommand{\eref}[1]{Eq.~(\ref{#1})}

\begin{document}

\title{
Non-equilibrium dynamics of drift–diffusion process under threshold resetting}

\author{Rahul Das}
\email{rahuldas@imsc.res.in}
\affiliation{The Institute of Mathematical Sciences, CIT Campus, Taramani, Chennai 600113, India \& Homi Bhabha National Institute, Training School Complex, Anushakti Nagar, Mumbai 400094, India}
\author{Satya N Majumdar}
\email{satyanarayan.majumdar@cnrs.fr}
\affiliation{Laboratoire de Physique Théorique et Modèles Statistiques (LPTMS),
Université de Paris-Sud, Bâtiment 100, 91405 Orsay Cedex, France}
\author{Arnab Pal}
\email{arnabpal@imsc.res.in}
\affiliation{The Institute of Mathematical Sciences, CIT Campus, Taramani, Chennai 600113, India \& Homi Bhabha National Institute, Training School Complex, Anushakti Nagar, Mumbai 400094, India}


\begin{abstract}

We study the emergence of non-equilibrium steady states (NESS) in stochastic processes under threshold resetting, an event-driven protocol in which the system resets to its initial configuration upon crossing a prescribed spatial boundary (threshold). In contrast to externally driven resetting, whose steady-state properties are well understood, the behavior under threshold resetting remains largely unexplored.
We derive general conditions for the existence of a NESS and show that, whenever it exists, the steady state at a given position $x$ can be expressed as the ratio of two fundamental quantities: the mean local time (MLT) at $x$ and the mean first-passage time (MFPT) to hit the threshold. In particular, for noisy systems, a finite MFPT guarantees the existence of a NESS. As an illustrative example, we analyze a drift–diffusion process in one dimension and uncover rich intermediate-time dynamics, including anomalous relaxation in the spatial distribution, damped oscillations in the moments and in the mean-squared displacement (MSD), governed by system parameters. Our findings provide a general understanding of the emergence of non-equilibrium steady states and relaxation dynamics under threshold resetting, revealing how threshold-induced events shape the spatial and temporal properties of a broad class of stochastic processes.

\end{abstract}

\maketitle


\section{Introduction}

Stochastic resetting has emerged as a vibrant area of research in statistical physics since the inception work \cite{evans_diffusion_2011} over a decade back. In its simplest form, resetting interrupts an ongoing stochastic process and returns it to a prescribed configuration at random times~\cite{evans_stochastic_2020,pal2015diffusion,pal2016diffusion, chechkin2018random, nagar2016diffusion, evans2018run}. Despite its conceptual simplicity, resetting gives rise to a wide range of nonequilibrium phenomena, including the emergence of nonequilibrium steady states~\cite{evans_stochastic_2020, pal2015diffusion,pal2016diffusion,majumdar2015dynamical}, optimization of first-passage and search processes~\cite{kusmierz2014first, pal2015diffusion,reuveni_optimal_2016,pal_first_2017, pal_first_2019, biswas2024search,pal2023thermodynamic,campos2015phase, pal2024random}, and anomalous relaxation and fluctuation properties~\cite{gupta2014fluctuating, eule2016non, gupta2020work, biroli2023extreme, gupta2022stochastic, pogorzelec2023resetting, sokolov2023linear, ghosh2023autonomous, mendez2016characterization, olsen2024thermodynamic}. Owing to its broad applicability, stochastic resetting has found connections across diverse disciplines including ecology~\cite{viswanathan1999optimizing, pal_search_2020,paramanick2024uncovering}, computer science~\cite{luby1993optimal, huang2007effect}, algorithm development \cite{blumer2024combining}, record statistics \cite{kumar2023universal} and operations research~\cite{bonomo2021mitigating, maurer2001restart,roy2024queues}. The subject has also found interesting applications and connections in chemistry and biophysics, including chemical reactions \cite{reuveni2014role,rotbart2015michaelis,biswas2023rate}, the anillin mechanism \cite{budnar2019anillin}, backtracking processes \cite{roldan2016stochastic}, facilitated diffusion \cite{rajoria2026broad}, and proofreading mechanisms \cite{biswas2026proofreading}. For a comprehensive overview of recent developments in the field, we refer the reader to the reviews \cite{evans_stochastic_2020,pal2022inspection,gupta2022stochastic,pal2024random,keidar2026stochastic}.



In the paradigmatic framework of stochastic resetting introduced in \cite{evans_diffusion_2011}, a stochastic process is restarted at random times determined by an external clock or distribution, independent of the underlying system dynamics. While such externally imposed resetting protocols have been extensively studied, an alternative and practically relevant mechanism arises through threshold-crossing events \cite{de2020optimization,de2021optimization,biswas2025target,biswas2026optimal}, where resetting is triggered whenever the system reaches a predefined threshold. Unlike externally timed resetting, the occurrence of reset events is therefore intrinsically coupled to the system dynamics itself.

Threshold-crossing events play an important role across a broad range of physical, biological, and engineered systems, often acting as operational limits or safety mechanisms that regulate system behavior. A prominent example is the integrate-and-fire neuron model, where the membrane potential evolves under external stimuli until it reaches a firing threshold, after which the neuron emits an action potential and resets to its resting state \cite{burkitt2006review,bachar2012stochastic,gerstein1964random}. In finance, stop-loss and take-profit strategies employ predefined thresholds to trigger asset transactions, thereby limiting losses or locking in gains \cite{kaufman2013trading,shiryaev2007optimal,zhang2001stock,miller1966model}. In physics, fibre bundle models describe the collective failure of materials in which fibres carry a common load until rupture thresholds are exceeded, leading to redistribution of stress among the surviving fibres \cite{pradhan2010failure,hansen2015fiber}. Similar threshold mechanisms appear in software engineering, where circuit breakers prevent systems from repeatedly accessing failing services, thereby improving stability and resilience \cite{nygard2018release,surendro2021circuit,montesi2018decorator}. Threshold-based control has also been used in chaotic systems, where simple stroboscopic mechanisms induce sustained temporal regularity with applications to laser dynamics \cite{sinha2001using,bhowmick2014targeting}. Recent works discussing temporal structure of the cell division \cite{kumar2026branching,corigliano2026essential} also hint toward an important role of threshold in cell biology, where division is triggered once a cell accumulates a critical amount of some internal variable. 

\begin{figure}[t]
    \centering
    \includegraphics[width=1\linewidth]{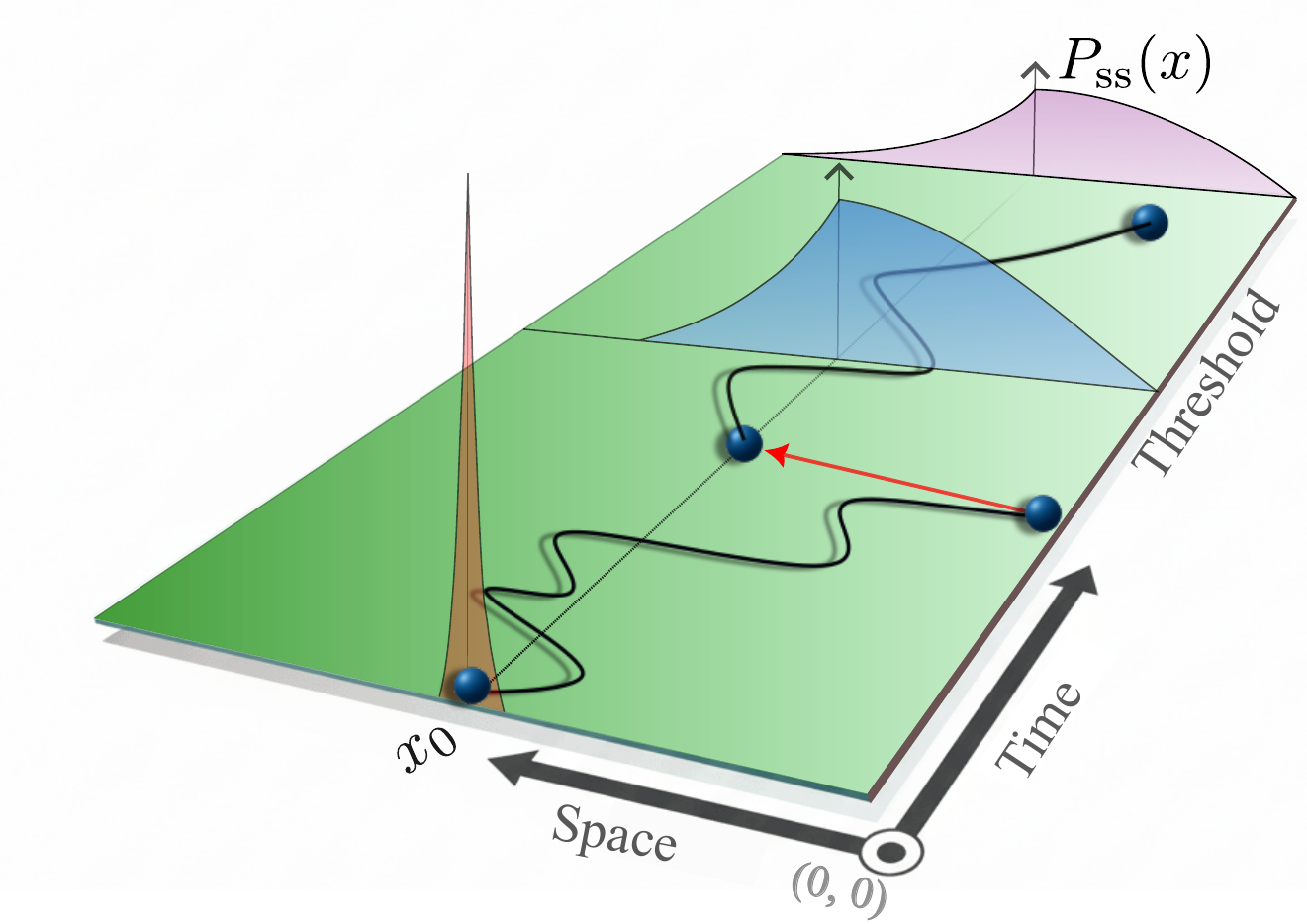}
    \caption{
    Schematic illustration of threshold resetting on the semi-infinite line.
A particle starts at $x=x_0$ and evolves under its intrinsic dynamics until it
reaches the threshold at the origin, where it is instantaneously reset to $x_0$.
Repeated threshold resetting events lead, at long times, to a stationary distribution
$P_{\mathrm{ss}}(x)$.}
    \label{fig:schematic}
\end{figure}

While threshold-driven processes have been extensively explored in various contexts, their implications for non-equilibrium properties and stochastic search dynamics remain relatively less understood. Initial studies on threshold resetting (TR) focused primarily on optimization aspects at the single-particle level. In particular, Refs.~\cite{de2020optimization,de2021optimization} investigated the efficiency and cost associated with a single diffusing particle subject to threshold resetting. More recently, the spatial properties of many-particle systems under TR have begun to attract attention. The steady-state structure, order statistics, and counting statistics of $N$ non-interacting diffusing particles evolving under threshold resetting were analyzed in \cite{biroli2026first}. In a different direction, Ref.~\cite{biswas2025target,biswas2026optimal} introduced a first-passage optimization framework for multiple searchers based on TR. Remarkably, although the searchers are intrinsically independent, simultaneous resetting events induced by threshold crossings generate effective correlations between them, thereby coupling their dynamics in a nontrivial manner \cite{biroli2023extreme,biroli2023critical,biroli2024exact,boyer2026emerging,de2026dynamically,biroli2024dynamically,biroli2025experimental,sabhapandit2024noninteracting,mesquita2025dynamically,galla2026diffusion,olsen2026information}. Such emergent collective behavior has recently received significant attention, including experimental demonstrations \cite{biroli2026experimental} and broader perspectives on dynamically induced correlations \cite{majumdar2026dynamically, mesquita2025dynamically2, de2026dynamically, boyer2026emerging, de_Mauro_2026}.

Despite these advances, several fundamental questions concerning the non-equilibrium nature of threshold resetting remain open. In conventional stochastic resetting, externally imposed reset events are known to generate non-equilibrium steady states (NESS) together with rich relaxation dynamics. Whether similar mechanisms arise when resetting is triggered internally through threshold crossings is still not fully understood (see Fig. (\ref{fig:schematic})). In particular, it remains unclear how threshold events give rise to stationary spatial structures, whether a general criterion exists for the emergence of NESS under TR, and what physical mechanisms govern their formation. Equally important is the characterization of the transient dynamics and the relaxation pathway toward the stationary state under threshold resetting. Addressing these questions is essential for developing a broader understanding of event-driven resetting processes and their non-equilibrium properties. In this work, we aim to address these issues and provide a systematic characterization of the emergence and relaxation of non-equilibrium steady states under threshold resetting.

For the convenience of the readers, we briefly summarize our main results here. Using a renewal formalism, we show that the existence of a non-equilibrium steady state is guaranteed whenever the mean first-passage time (MFPT) to the threshold boundary is finite. In this regime, the steady-state distribution acquires a simple and physically transparent representation: it can be expressed as the ratio between the mean local time (MLT) and the MFPT associated with reaching the threshold. 
To demonstrate these features, we consider a drift–diffusion (DD) process, previously studied extensively in the presence of stochastic resetting \cite{ray2019peclet,pal2019local}. Here, we revisit this process in the context of threshold resetting.  Beyond the stationary behavior, we also investigate the intermediate-time dynamics of spatial observables, such as the mean position and the mean-square displacement (MSD), and uncover the emergence of anomalous and oscillatory behavior during the relaxation process. 

This article is organized as follows.
In Sec.~\ref{general_fw_section}, we present the renewal formalism, following \cite{de2020optimization}, for a stochastic dynamics under threshold resetting (TR) and derive the steady-state distribution under TR and express the same in terms of two physical quantities: the MLT and the MFPT. We discuss a sufficient condition for the emergence of the steady state.
In Sec.~\ref{dd_section}, we analyze the drift–diffusion process under threshold resetting, focusing on the transient and steady state properties of the propagator and moments. In Sec.~\ref{sec:DT}, we show a dynamical phase transition of a spatial point from a transient state to a steady state at long times. We summarize our main findings and conclude with future directions in Sec. \ref{conclusion}. Supporting derivations and results are presented in the Appendix.

\section{Set-up and Framework}\label{general_fw_section}
Let us consider a particle moving in one dimension that starts its motion from $x=x_0$, with dynamics governed by an arbitrary stochastic process. During its evolution, whenever the particle reaches the origin, $x=0$, which we identify as the \textit{threshold}, it is instantaneously reset to its initial position $x=x_0$. This mechanism is referred to as \textit{threshold resetting} (TR). Following each reset event, the particle resumes its dynamics anew from the initial state. Our main quantity of interest is the probability density of finding the particle at position $x$ at time $t$, denoted by $\mathbb{P}(x,t|x_0)$. An exact formal expression for $\mathbb{P}(x,t|x_0)$ under TR was derived in \cite{de2020optimization} using a renewal formalism, which we briefly review here for completeness. Building upon this framework, we then explicitly characterize the steady states emerging from the TR mechanism and analyze the intermediate-time behavior before the system reaches these stationary states.

\subsection{The renewal formalism}\label{renewal_form_section}
To obtain the propagator $\mathbb{P}(x, t|x_0)$, we first note that it has a contribution from two types of trajectories - i) those that did not undergo any resetting events till time $t$, and ii) those which had undergone one or multiple resetting events before $t$. Taking into account both of these events, one can write a simple renewal equation for the propagator as follows \cite{de2020optimization}
\begin{equation}
    \mathbb{P}(x, t|x_0) = \mathbb{G}_{+}(x, t |x_0) + \int_{0}^{t}dt^\prime ~\mathbb{F}(t^\prime| x_0)\mathbb{P}(x, t - t^\prime |x_0), \\
    \label{renewal}
\end{equation}
where $\mathbb{G}_{+}(x, t |x_0)$ is the propagator in the presence of an absorbing boundary at the origin (here, $+$ stands as an indicator that the particle stays in the positive half-line) and $\mathbb{F}(t|x_0)$ is the first-passage time density to reach the threshold starting from $x_0$. 
One can derive the latter from the former quantity. Integrating the restricted propagator $\mathbb{G}_{+}(x, t|x_0)$ over the final position $x$, gives the survival probability, namely
\begin{equation}\label{surv_def}
\mathbb{S}(t|x_0) = \int_0^{\infty} \mathbb{G}_{+}(x,t|x_0) dx,
\end{equation}
and the first-passage probability density is then simply $\mathbb{F}(t|x_0)= - \partial_t \mathbb{S}(t|x_0)$. Consequently, the mean first-passage time (MFPT) is given by
\begin{align}\label{mfpt_def}
\langle T \rangle = \int_0^{\infty} t~ \mathbb{F}(t|x_0) dt &= \int_0^{\infty} \mathbb{S}(t|x_0) dt \nonumber \\
&= \int_0^{\infty} dx \int_0^{\infty} \mathbb{G}_{+}(x,t|x_0) dt.
\end{align}

The renewal equation \eref{renewal} can be physically understood in the following way: the first term in the RHS accounts for the trajectories which have encountered no resetting events in between $[0,t]$. The probability of the same is given by the reset-free propagator $\mathbb{G}_{+}(x, t |x_0)$ with the threshold at the origin acting as an absorbing boundary since the particle stops its motion exactly there and can not continue further. On the other hand, the second term takes care of the trajectories which had at least one resetting event before $t$. The probability that the first threshold resetting event occurs between $[t^\prime,t^\prime+dt^\prime]$ is the same as the probability when the particle makes a first passage to the threshold target between that time interval, which is given by $dt^\prime ~\mathbb{F}(t^\prime|x_0)$. Furthermore, after the first resetting event, the particle starts its motion anew from the initial position $x_0$ for the remaining duration $t-t^\prime$, accounting for $\mathbb{P}(x,t-t^\prime|x_0)$. Finally, noting that the first resetting can happen anywhere between $[0,t]$, we integrate the combined probability  $dt^\prime ~\mathbb{F}(t^\prime|x_0)\mathbb{P}(x, t - t^\prime |x_0)$ to reach the second term in \eref{renewal}.

Performing a Laplace transformation on the renewal equation \eref{renewal} immediately yields the following expression \cite{de2020optimization}
\begin{align}
    \widetilde{\mathbb{P}}(x, s|x_0) &= \frac{\widetilde{\mathbb{G}}_{+}(x, s|x_0)}{1 - \widetilde{\mathbb{F}}(s|x_0)}
    \label{pdf-ls},
\end{align}
where $\widetilde{f}(s)=\int_0^\infty dt e^{-st}f(t)$ denotes Laplace transform for any function $f(t)$. 
The denominator in Eq. (4) can also be expressed in terms of the Laplace transform $\widetilde{\mathbb{G}}_+(x,s\vert x_0)$.
Taking the Laplace transform of the relation $\mathbb{F}(t|x_0)= - \partial_t \mathbb{S}(t|x_0)$ with respect to $t$ and using $\mathbb{S}(0\vert{}x_0)=1$, one obtains
\begin{equation}\label{eqA}
\widetilde{\mathbb{F}}(s|x_0)= 1- s ~\widetilde{\mathbb{S}}(s|x_0) ,
\end{equation}
where $\widetilde{\mathbb{S}}(s\vert{}x_0)$ is the Laplace transform of the survival probability $\mathbb{S}(t\vert{}x_0)$. Further, using Eq. \eqref{surv_def} gives
\begin{equation}\label{eqB}
\widetilde{\mathbb{F}}(s|x_0)= 1- s \int_0^{\infty} dx \, \widetilde{\mathbb{G}}_+ (x,s|x_0). 
\end{equation}
Consequently, Eq. \eqref{pdf-ls} becomes
\begin{equation}
\widetilde{\mathbb{P}}(x,s|x_0)= \frac{ \widetilde{\mathbb{G}}_+(x,s|x_0)}{ s \int_0^{\infty} \widetilde{\mathbb{G}}_+(x,s|x_0) \, dx} \label{eqC}.
\end{equation}
Thus, the propagator under TR is expressed, in the Laplace space, entirely in terms of the constrained propagator $\widetilde{\mathbb{G}}_+(x,s\vert{}x_0)$ of the corresponding reset-free dynamics. This result is particularly useful, as it implies that knowledge of the underlying dynamics in the absence of resetting is sufficient to determine the propagator under TR, akin to the stochastic resetting paradigm \cite{evans_diffusion_2011,evans_stochastic_2020}. Since the quantity $\widetilde{\mathbb{G}}_{+}(x, s|x_0)$ is exactly known for a variety of physical problems, this representation makes the analysis of the propagator under TR analytically tractable to a great extent. We will use these results to investigate the steady states induced by the TR mechanism.

\subsection{Steady state under TR}\label{sufficient_condition_Formalism}
If a steady-state (SS) exists, $\mathbb{P}(x, t|x_0)$ becomes time-independent at long times, i.e., $\mathbb{P}(x, t \to \infty|x_0) = \mathbb{P}_{\text{ss}}(x)$. Consequently, its Laplace transform with respect to $t$ behaves for small $s$ as
\begin{align}
\widetilde{\mathbb{P}}(x,s|x_0)=& \int_0^{\infty} \mathbb{P}(x,t|x_0) e^{-st} dt  \quad \\ &\to \frac{1}{s} \mathbb{P}_{\text{ss}}(x)  ~~\text{as}~ s\to 0. \nonumber
\end{align}
This gives rise to the following relation (also known as the final value theorem)

\begin{align}
    \mathbb{P}_{\text{ss}}(x) &= \lim_{s \to 0} s \widetilde{\mathbb{P}}(x, s|x_0) = \lim_{s \to 0} \frac{s \widetilde{\mathbb{G}}_{+}(x, s|x_0)}{1 - \widetilde{\mathbb{F}}(s|x_0)} .
    \label{pdf-ss-1}
\end{align}
If the particle has a \textit{finite} mean first-passage time (MFPT) to the threshold, denoted by $\langle T \rangle$ with $T$ being the random first-passage time, then the denominator in \eref{pdf-ss-1} can be expanded up to first order in $s$ to have
\begin{align}
    \lim_{s\to 0}\widetilde{\mathbb{F}}(s|x_0) \approx 1- s\langle T \rangle + o(s).
\end{align}
Note that the above relation simply follows from Eq. \eqref{mfpt_def} and \eqref{eqA} in the small $s$ limit. Plugging this expression back to \eref{pdf-ss-1} then yields
\begin{align}
    \mathbb{P}_{\text{ss}}(x) &= \frac{\int_0^\infty dt ~\mathbb{G}_{+}(x,t|x_0)}{\langle T \rangle}, \label{pdf-ss}
\end{align}
where we have used the fact that $    \lim_{s\to 0}\widetilde{\mathbb{G}}_{+}(x, s|x_0) = \int_0^\infty dt ~\mathbb{G}_{+}(x,t|x_0)$. 
The above result had appeared earlier in \cite{biroli2026first} for a general case of $N$ diffusive particles under TR, which subsequently can be extended to any underlying dynamics. The single particle case $(N = 1)$ converges to the above-mentioned equation. Looking closely, the form of the steady state has an interesting physical interpretation, which we explain below.

The quantity $\int_0^\infty dt ~\mathbb{G}_{+}(x,t|x_0)$ can be interpreted as the mean local time (MLT) of the particle at a given position $x$ during a single excursion between two successive threshold crossings.
To see this, let us recall that the local time $L(x|x_0)$ spent at position $x$ starting from $x = x_0$ along a single trajectory $X(t)$ is defined as follows
\begin{align}
    L(x|x_0) = \int_0^\infty dt \, \delta\!\left[X(t) - x\right].
\end{align}
To obtain the MLT, we need to average this over many realizations of $X(t)$ weighted with $\mathbb{G}_{+}(X, t|x_0)$ so that
\begin{align}\label{MLT_prop}
    \langle L(x|x_0) \rangle 
    &= \int_0^\infty dt\, \langle \delta\!\left[X(t) - x\right] \rangle \nonumber \\
    &= \int_0^\infty dt \int_0^{\infty} \!dX\, \delta\!\left[X(t) - x\right]\, \mathbb{G}_{+}(X, t | x_0) \nonumber \\
    &= \int_0^\infty dt\, \mathbb{G}_{+}(x, t | x_0).
\end{align}
By replacing the above relation into Eq. \eqref{pdf-ss}, we get 
\begin{equation}\label{pdf-ratio}
    \mathbb{P}_{\text{ss}}(x) = \frac{\langle L(x| x_0)\rangle}{\langle T \rangle},
\end{equation}
which can be interpreted in the following way 
\begin{equation}
    \mathbb{P}_{\text{ss}}(x) = \frac{\text{average time spent at $x$ in one cycle}}{\text{average duration of one cycle}},
\end{equation}
where a \textit{cycle} represents the event between two threshold driven first-passage occurrence.
It follows from Eq. \eqref{mfpt_def} and \eqref{MLT_prop} that
\begin{equation}\label{mfpt_mlt_reln}
    \langle T\rangle = \int_0^\infty dx \langle L(x|x_0)\rangle.
\end{equation}
which also shows that $\mathbb{P}_{\text{ss}}(x)$ in Eq. \eqref{pdf-ss}, when it exists, is normalized to unity, i.e., $\int_0^\infty dx~\mathbb{P}_{\text{ss}}(x) = 1$.

The relation in Eq. \eqref{mfpt_mlt_reln} implies that, if the MFPT is finite, then the MLT must fall into one of the following two categories:
\begin{itemize}
    \item $\langle L(x|x_0)\rangle$ remains finite for all $x$, so that the integral over the entire domain converges.
    \item $\langle L(x|x_0)\rangle$ exhibits integrable singularities at certain points in $x$, meaning that the MLT diverges locally but the divergence is sufficiently weak for the integral over $x$ to remain finite, ensuring a finite MFPT.
\end{itemize}
In the former case, one obtains a continuous steady-state distribution over space, whereas in the latter case, the steady-state distribution has singularities as it will diverge at some spatial points, which can be seen from \eref{pdf-ratio}. This result establishes that a finite MFPT to reach the threshold provides a sufficient criterion for the existence of a steady state under TR in a noisy system.
In the following section, we consider drift-diffusion under TR to delve deeper to understand the steady-state and transient properties of the particle.


\section{Drift-diffusion under TR}\label{dd_section}
Let us now consider a drift–diffusion (DD) process where a constant drift $v>0$ is directed toward the threshold. This ensures a finite mean first-passage time (MFPT) to find the threshold boundary and subsequently renders a steady state under TR. Another simple system would be $N (>2)$ diffusive particles, which ensures a finite MFPT and consequently a steady state under TR without any drift. This was studied in \cite{biroli2026first}. In this study, we examine single-particle drift-diffusion and its properties. The propagator $\mathbb{G}_{+}(x,t|x_0)$ satisfies the following Fokker-Planck equation \cite{redner2001}
\begin{equation}\label{fokkerPlanck}
    \frac{\partial \mathbb{G}_{+}(x,t|x_0)}{\partial t}
    =
    v\,\frac{\partial \mathbb{G}_{+}(x,t|x_0)}{\partial x}
    +
    D\,\frac{\partial^2 \mathbb{G}_{+}(x,t|x_0)}{\partial x^2},
\end{equation}
subject to the initial and boundary conditions
$\mathbb{G}_{+}(x,0|x_0)=\delta(x-x_0)$ and $\mathbb{G}_{+}(0,t|x_0)=0, \mathbb{G}_{+}(x \to \infty, t|x_0) = 0$.
Here, $v$ denotes the drift velocity toward the origin and $D$ is the diffusion coefficient. 
It is quite convenient to introduce the dimensionless P\'eclet number given by
\begin{equation}\label{dimensionless}
    \text{Pe} = \frac{x_0 v}{2D},
\end{equation}
which quantifies the ratio between the timescales of advective and diffusive transport \cite{redner2001}. By solving Eq. \eqref{fokkerPlanck} in Laplace space, we find the propagator as
\begin{equation}\label{g+laplace_compact}
\begin{split}
    \widetilde{\mathbb{G}}_{+}(x,s|x_0) &= \frac{1}{v\lambda(s)} e^{\mathrm{Pe}(1- x/x_0)} \\
    &\quad \times \Big\{ e^{-\lambda(s)\mathrm{Pe}|x/x_0 - 1|} - e^{-\lambda(s)\mathrm{Pe}(x/x_0 + 1)} \Big\},
\end{split}
\end{equation}
where $\lambda(s)$ is a function of the Laplace parameter $s$, defined as
\begin{equation}\label{lambda_s}
    \lambda(s)=\sqrt{1+\frac{4Ds}{v^2}}.
\end{equation}
We now use Eqs. \eqref{eqB} and \eqref{g+laplace_compact}, to obtain the corresponding first-passage time density in Laplace space, which is given by \cite{redner2001}
\begin{equation}\label{fpLaplace}
    \widetilde{\mathbb{F}}(s|x_0)
    =
    \exp\!\left[
    \text{Pe}(1 - \lambda(s))
    \right].
\end{equation}
Substituting into the general renewal expression Eq. \eqref{pdf-ls}, we obtain the full propagator of the DD process under TR in Laplace space as the following
\begin{align}\label{exactPropLS}
    \widetilde{\mathbb{P}}(x, s|x_0) &= \frac{e^{\mathrm{Pe}(1- x/x_0)}}{v\lambda(s)[1 - e^{
    \text{Pe}(1 - \lambda(s))}]} \nonumber \\
    &\quad \times \Big\{ e^{-\lambda(s)\mathrm{Pe}|x/x_0 - 1|} - e^{-\lambda(s)\mathrm{Pe}(x/x_0 + 1)} \Big\}.
\end{align}
Using Laplace transform tables \cite{bateman_1954_mhd23-e0z22}, we can compute the inverse Laplace transform in terms of an infinite sum as follows
\begin{equation}\label{DD_propT}
\begin{split}
&\mathbb{P}(x,t\,|\,x_0) ={} \frac{e^{-v^2 t/4D}}{\sqrt{4\pi D t}}
    \sum_{n=0}^{\infty} e^{\mathrm{Pe}\,(n+1-x/x_0)} \\
& \qquad\times \Bigl[\, e^{-(n+|x/x_0-1|)^2 x_0^2/4Dt} - e^{-(n+x/x_0+1)^2 x_0^2/4Dt} \,\Bigr].
\end{split}
\end{equation}
The steady state and the approach towards this time-independent state are discussed in the following sub-sections. Notably, when $v < 0$, there is a finite probability that the particle escapes to infinity during any excursion. Consequently, the threshold resetting mechanism fails to confine the probability density, and the non-equilibrium steady state no longer exists.

\begin{figure*}
    \includegraphics[width=0.9\textwidth]{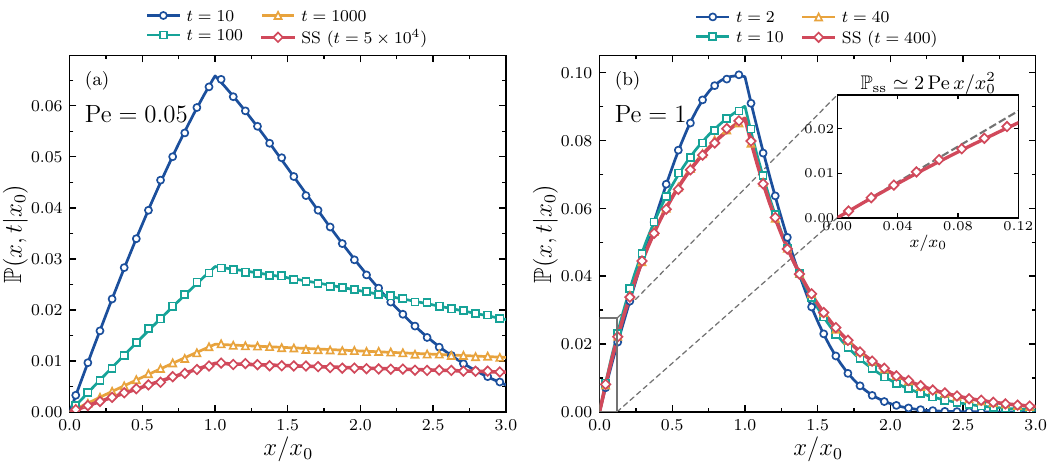}
    \caption{
Temporal evolution of the probability density $\mathbb{P}(x, t | x_0)$ for a particle undergoing drift-diffusion (DD) subject to threshold resetting (TR). The panels illustrate two distinct $\text{Pe}$ regimes: (a) diffusion-dominated regime ($\text{Pe} = 0.05$); (b) intermediate crossover regime ($\text{Pe} = 1$); the high \text{Pe} regime is discussed separately in Fig. \ref{Fig_HighPePropEvo}. The dynamic approach to the non-equilibrium steady state (SS, solid red line) is shown for different times. Solid lines represent exact analytical solutions, whereas markers denote Monte Carlo simulation data. In both panels, the kink at $x = x_0$ in the SS is clearly visible, as shown in Eq. \eqref{eq:jumpFirstDerivative} in terms of the discontinuity in the first spatial derivative. The inset in panel (b) highlights the linear behavior of the steady state, i.e.,  $\mathcal{P}_{\text{ss}}(u) \sim u$ (or $\mathbb{P}_\text{ss} \sim x$) near the threshold at the origin, valid for any $\text{Pe}$ regime. System parameters: (a) $x_0 = 10, v = 0.05, D = 5$; and (b) $x_0 = 10, v = 1, D = 5$.}
    \label{DD_prop_plot}
\end{figure*}

\subsection{Steady-state behavior}

\begin{figure}
    \centering
    \includegraphics[width=\linewidth]{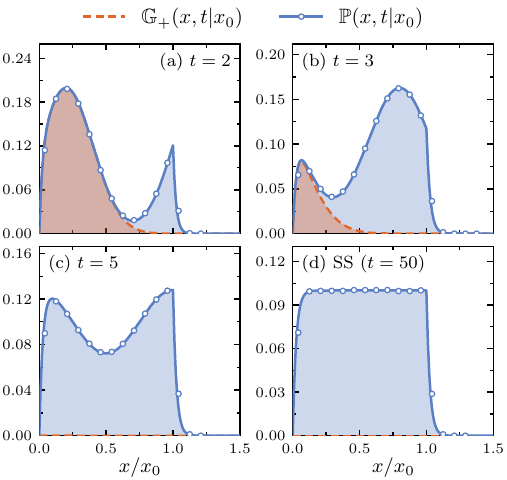}
    \caption{Temporal evolution of $\mathbb{P}(x,t |x_0)$, and $\mathbb{G}_{+}(x, t|x_0)$ for $\text{Pe} = 20$, or drift-dominated regime. Each panel shows the propagator at different times. Comparing the above figure with Eq. \eqref{PropRecur}, we find that a peak near the origin at early times ($ t = 2, 3$) arises due to the first (reset-free) term $\mathbb{G}_{+}(x, t|x_0)$. At later times, the reset-free contribution becomes very small, and eventually it relaxes to the steady state (SS) due to multiple resetting events. The kink in the SS at $x = x_0$ is visible in panel (d). The solid line is the analytical solution, and markers denote the simulation data. System parameters are: $x_0 = 10, v = 4, D = 1$.}
    \label{Fig_HighPePropEvo}
\end{figure}

In the long-time limit, the system relaxes to a stationary state. This can be computed simply by using the small $s$ (or long $t$) behavior of $\widetilde{\mathbb{P}}(x, s|x_0)$ (as in Eq. \eqref{pdf-ss-1}).

\begin{align}\label{SS}
    \mathbb{P}_{\mathrm{ss}}(x)= \frac{1}{x_0} \mathcal{P}_\text{ss}\left(u = \frac{x}{x_0} \right),
\end{align}
where $\mathcal{P}_\text{ss}(u)$ is defined in this way
\begin{equation}\label{DD_ss}
    \mathcal{P}_\text{ss}(u) =
    \begin{cases}
        1 - \exp[-2\text{Pe}\,u], & u \le 1 \\[6pt]
        \bigl(1 - \exp[-2\text{Pe}]\bigr) \exp[-2\text{Pe}(u-1)], & u > 1
    \end{cases}
\end{equation}
We have plotted Eq. (\ref{SS}) against numerical simulations for different Pe values in Fig.[\ref{DD_prop_plot}, \ref{Fig_HighPePropEvo}(d)], showing an excellent match. 

\begin{figure}[h]
    \centering
    \includegraphics[width=1\linewidth]{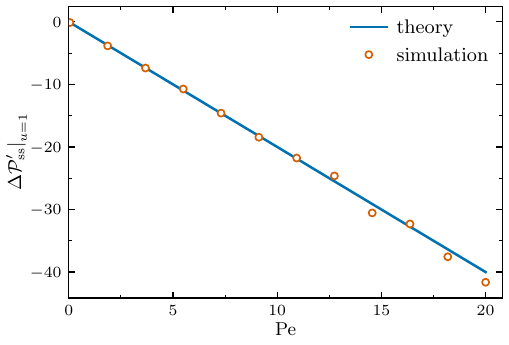}
    \caption{Verification of Eq.  \eqref{eq:jumpFirstDerivative}, regarding the jump discontinuity in the first derivative of the steady state at $u=1$, is shown for various $\text{Pe}$ with a slope of $-2$. Here, we change  velocity $v$, keeping $x_0$ and $D$ fixed. System parameters: $x_0 = 1, D = 0.5, v \in [0, 20]$.}
    \label{fig:kink}
\end{figure}

Several comments are in order. Since the particles are immediately reset to the initial position $u=1$ upon reaching the absorbing threshold at $u=0$, the steady-state probability density vanishes at the origin, i.e., $\mathcal{P}_{\mathrm{ss}}(0) = 0$. Close to this threshold boundary, the distribution grows linearly, $\mathcal{P}_{\text{ss}}(u) \propto u$, shown in Fig. [\ref{DD_prop_plot}(b) inset]. Conversely, in the region beyond the resetting position ($u > 1$), the distribution decays exponentially due to the finite drift driving the system back toward the threshold. The steepness of this exponential tail is governed by the P\'eclet number. Interesting shapes for the steady state emerge in the limits: diffusion-dominated ($\text{Pe} \ll 1$) or drift-dominated ($\text{Pe} \gg 1$). For $\text{Pe} \ll 1$, $\mathcal{P}_{\mathrm{ss}}(u)$ increases linearly up to the reset point $u = 1$, beyond which it exhibits a slow spatial decay (see Fig. \ref{DD_prop_plot}(a)). In the strong drift limit ($\text{Pe} \gg 1$), the propagator develops a peak in the immediate vicinity of the threshold, then redistributes to rapidly flatten out to a constant plateau, i.e., $\mathcal{P}_{\mathrm{ss}}(u) \approx 1$ across the $0 < u \le 1$ domain, followed by a sharp exponential truncation for $u > 1$. These distinct spatial profiles are further studied and illustrated in Fig.~\ref{Fig_HighPePropEvo}. An interesting feature of the steady-state distribution valid for any $\text{Pe}$ regime is the presence of a kink at $u = 1$, which means the first derivative of $\mathcal{P}_{\text{ss}}(u)$ with respect to $u$ is discontinuous at $u =1$. This follows from Eq. \eqref{DD_ss}  that
\begin{equation}\label{eq:jumpFirstDerivative}
    \Delta \mathcal{P}^\prime_{\text{ss}}\big|_{u = 1} := \mathcal{P}^\prime_{\text{ss}}(u \to 1+) - \mathcal{P}^\prime_{\text{ss}}(u \to 1-) = -2 \text{Pe},
\end{equation}
where $\mathcal{P}^\prime_{\text{ss}}(u) = d\mathcal{P}_{\text{ss}}(u)/du$. The presence of this kink in the steady state is clearly visible in Fig. \ref{DD_prop_plot} and Fig. \ref{Fig_HighPePropEvo}(d), and a verification of the above result against numerical simulation is presented in Fig. \ref{fig:kink}.

Quite interestingly, the transient relaxation toward these steady states differs markedly between the regimes. Most notably, in the drift-dominated limit ($\text{Pe} \gg 1$), the propagator $\mathbb{P}(x,t|x_0)$ develops a bimodal (two-peak) profile at intermediate times [see Fig.~\ref{Fig_HighPePropEvo}]. This behavior can be physically understood by rewriting Eq. (\ref{renewal}) in the following way 
\begin{align}\label{PropRecur}
    \mathbb{P}(x, t|x_0) = & \mathbb{G}_{+}(x,t|x_0) + \nonumber \\
    &\sum_{n \ge 1} \int_0^t dt^\prime \, \mathbb{F}_n(t^\prime|x_0)\mathbb{G}_{+}(x, t - t^\prime|x_0),
\end{align}
where $\mathbb{F}_n(t|x_0)$ is the first-passage time distribution for reaching the threshold for the $n$-th time (so that $\mathbb{F}_1 \equiv \mathbb{F}$). Recall that the first term, $\mathbb{G}_{+}(x,t|x_0)$, represents the survival propagator in the absence of resetting; driven by the strong drift, the surviving ensemble manifests as a moving probability packet traveling toward the threshold at $x=0$. At the initial times, this packet gives rise to one of the peaks in the propagator near the origin. It is shown in Fig. \ref{Fig_HighPePropEvo}(a). This is the no-resetting case since the envelope of $\mathbb{G}_{+}(x,t|x_0)$ (shown by the dashed line) significantly overlaps with the full PDF. Upon a resetting event, the particle is immediately pushed back toward $x_0$, rendering the probability to locally accumulate at the resetting position, forming the second peak -- see Fig. \ref{Fig_HighPePropEvo}(b). Following this, there are effects due to multiple resetting events. Evidently, as time grows, the contribution due to $\mathbb{G}_{+}(x, t|x_0)$ becomes negligible, and we see redistribution of the probability near $x = x_0$ and the threshold -- see Fig. \ref{Fig_HighPePropEvo}(c). At long time, the particle can be found with equal probability in the domain $x \in (0,x_0)$, rendering a plateau-like steady state -- see Fig. \ref{Fig_HighPePropEvo}(d).

\subsection{Mean and mean squared displacement}
\begin{figure}
    \centering
    \includegraphics[width=\linewidth]{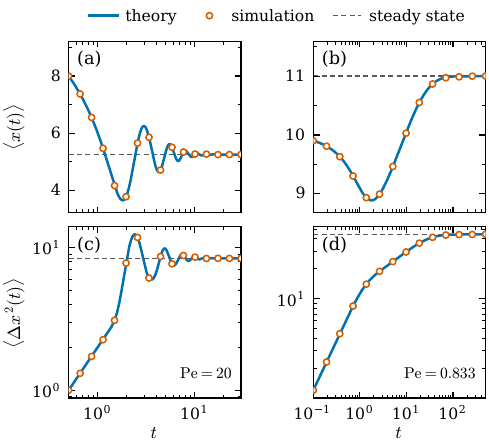}
    \caption{Temporal evolution of Mean Square Displacement (MSD) and mean position for drift-diffusion under threshold resetting (TR). Panels (a) and (c) illustrate the oscillatory regime for mean position and MSD, respectively, using parameters $x_0 = 10, v = 4, D = 1$, i.e., $\text{Pe} = 20$. Panels (b) and (d) show the transition to non-oscillatory behavior when diffusion dominates, or drift is reduced. Parameters used: $x_0 = 10, v = 1, D = 6$, i.e., $\text{Pe} = 0.833$. Solid lines represent the theory, while colored points denote simulation results.}
    \label{fig:Osc_DD}
\end{figure}

We now turn our attention to analyze the mean position and MSD of the particle under TR. Using the propagator of the particle in time, or in the Laplace domain, one can compute the $k$-th moment of $x$ in their respective domains as follows.
\begin{align}
    \langle x^k(t)\rangle &= \int_0^\infty x^k ~\mathbb{P}(x, t|x_0) ~dx,\\
    \langle \widetilde{x^k}(s)\rangle &= \int_0^\infty x^k ~\widetilde{\mathbb{P}}(x, s|x_0) ~dx.
\end{align}
Using Eq. \eqref{exactPropLS}, the exact results for the first two moments in Laplace space for our drift-diffusion under the TR case read
\begin{align}
\label{DD_mean_ls}
\langle \widetilde{x}(s)\rangle
&=
\frac{1}{s}\frac{x_0}{1-\exp\!\left[\text{Pe}(1-\lambda(s))\right]}
-
\frac{v}{s^2},\\
\langle \widetilde{x^2}(s)\rangle &= \frac{2 v^2}{s^3} + \frac{2 D}{s^2} \nonumber\\
&\quad + \frac{1}{1-\exp\!\left[\text{Pe}(1-\lambda(s))\right]}\left(\frac{x_0^2}{s}-\frac{2 x_0 v}{s^2} \right),
\end{align}
which can be inverted exactly using table \cite{bateman_1954_mhd23-e0z22}. We can express them in summation form as below.

\begin{align}
\langle x(t)\rangle &= x_0 - vt + \frac{x_0}{2}\sum_{n\ge 1}\mathcal{A}_n(t),
  \label{dd_il_table}\\
\langle x^2(t)\rangle &= 2Dt + (x_0 - vt)^2
  + x_0^2 \sum_{n\ge 1}\Bigl[\tfrac12\mathcal{A}_n(t) - \mathcal{B}_n(t)\Bigr],
  \label{dd_il_table2}
\end{align}
where the dimensionless functions $\mathcal{A}_n(t)$ and $\mathcal{B}_n(t)$ are defined as

\begin{subequations}\label{dd_AB_defs}
\begin{align}
\mathcal{A}_n(t) &= \operatorname{erfc}\bigl[\xi_n^-(t)\bigr]
  + e^{2n\mathrm{Pe}}\operatorname{erfc}\bigl[\xi_n^+(t)\bigr],\\
\mathcal{B}_n(t) &= \Bigl(n+\frac{vt}{x_0}\Bigr)e^{2n\mathrm{Pe}}
  \operatorname{erfc}\bigl[\xi_n^+(t)\bigr]\nonumber\\
&\quad - \Bigl(n-\frac{vt}{x_0}\Bigr)\operatorname{erfc}\bigl[\xi_n^-(t)\bigr],\\
\xi_n^\pm(t) &= \frac{n x_0 \pm vt}{\sqrt{4Dt}}.
\end{align}
\end{subequations}
The MSD can be found using the relation, $\langle \Delta x^2(t) \rangle = \langle x^2(t)\rangle-\langle x(t)\rangle^2$. At long times, both these quantities converge to steady-state values, which can be computed easily using the small-$s$ behavior of their Laplace transforms (namely, the final value theorem)
\begin{align}
    \langle x\rangle_{\text{ss}}
    &=
    \frac{x_0}{2}\!\left(1+\frac{1}{\text{Pe}}\right),\label{ss_mean_msd_dd}\\
    \langle \Delta x^2 \rangle_{\text{ss}}
    &=
    \frac{x_0^2}{12}\!\left(1+\frac{3}{\text{Pe}^2}\right).
\end{align}
The temporal behavior (along with their respective steady-state convergence at long times) of these quantities are shown in Fig. (\ref{fig:Osc_DD}) for different values of Pe. Quite interestingly, we observe that in some cases, the moments have non-monotonic intermediate time behavior (oscillations) before converging to the steady state. We discuss this next. 

Although the exact time-dependent expressions capture the full dynamics, the origin of the oscillatory behavior is not immediately apparent. To elucidate this, we analyze the inverse Laplace representation of the first moment using the Bromwich integral
\begin{equation}
\label{DD_bromwich}
\langle x(t)\rangle
=
\frac{1}{2\pi i}
\int_{c-i\infty}^{c+i\infty}
ds\,e^{st}\langle \widetilde{x}(s)\rangle,
\end{equation}
where the Bromwich contour is chosen such that all singularities of $e^{st}\langle \widetilde{x}(s)\rangle$ lie to the left of $\mathrm{Re}(s)=c$ and the exact expression of $\langle \widetilde{x}(s)\rangle$ is given in Eq. \eqref{DD_mean_ls}. The function $\langle \widetilde{x}(s)\rangle$ consists of a branch point at
\begin{equation}\label{BP_main}
s_b = -\frac{v^2}{4D},
\end{equation}
and a discrete set of poles,
\begin{align}
s_n &= - \gamma_n - i\,{\omega}_n,\\
    &= -\frac{4\pi^2 n^2 D}{x_0^2} - i \frac{2\pi n v}{x_0}, \qquad n \in \mathbb{Z}.
\end{align}
Evidently, the pole at $s=0$ (corresponding to $n=0$) determines the stationary state, while the presence of other poles (complex in nature) gives rise to oscillatory contributions in the intermediate-time regime, while their negative real parts ensure exponential damping. The mean position can thus be expressed as
\begin{equation}\label{DD_mean_time}
\begin{split}
\langle x(t)\rangle = \langle x\rangle_{\text{ss}}
+ 
\sum_{n=1}^\infty
\mathcal{C}_n \cos(\omega_n t+\theta_n)e^{-\gamma_n t}
- \mathcal{H}(t).
\end{split}
\end{equation}
where the second term is the oscillatory contribution associated with the poles, and $\mathcal{H}(t)$ represents the contribution from the branch cut. Explicit expressions for the dimensionless amplitudes $\mathcal{C}_n$, the phases $\theta_n$ and the function $\mathcal{H}(t)$ are provided in Appendix~\ref{DD_appdx}. This form demonstrates exponential relaxation toward the stationary state, modulated by damped oscillations.

Crucially, 
the emergence of oscillations depends on the competition between oscillatory and damping parts. This can be shown using the ratio between the absolute values of the real and imaginary parts of the complex poles, which is
\begin{equation}
    \frac{\omega_n}{\gamma_n} = \frac{x_0 v}{2 \pi n D} = \frac{\text{Pe}}{\pi n}.
\end{equation}
Clearly, it depends on the system parameters—the distance between the initial position and the threshold $x_0$, drift velocity $v$, and diffusion coefficient $D$. As introduced in Eq. \eqref{dimensionless}, these combine into the P\'eclet number Pe. A high P\'eclet number ($\text{Pe} \gg 1$) implies advection dominance, where the particle reaches the threshold within a narrow temporal window centered around the MFPT $\langle T \rangle = x_0/v$ and then returns to the resetting position. Thus, the oscillations in the drift-dominated regime arise from the near-synchronization of renewal events induced by threshold resetting when resets occur in coherent bursts. This repeated cycle of drift, hitting the threshold, and returning to the initial position $x_0$ generates periodic probability waves (as seen in Fig. \ref{Fig_HighPePropEvo}), leading to oscillations in the moments [Fig.~\ref{fig:Osc_DD}(a, c)]. For a smaller Pe value, diffusion will gradually broaden the first-passage time distribution, thereby destroying the synchronization between renewal events and causing the oscillations to become completely suppressed and an exponential convergence to the steady state values is observed. 

\section{Dynamical transition from transient to steady-state}\label{sec:DT}

\begin{figure}
    \centering
    \includegraphics[width=1\linewidth]{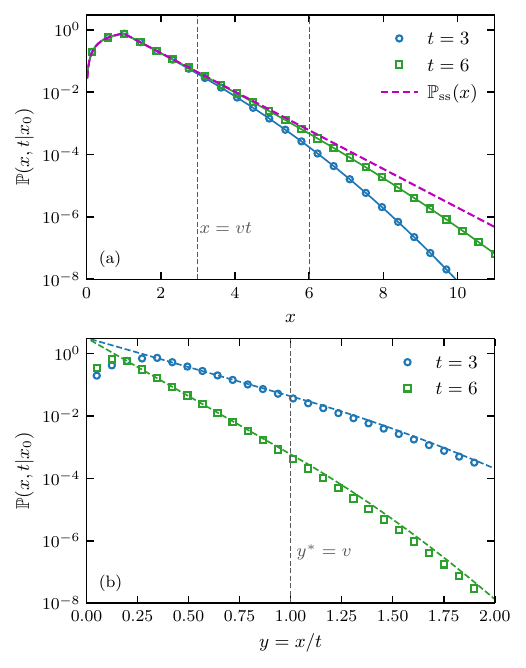}
    \caption{The propagator of a drift-diffusing particle under threshold resetting. Panel (a): The symbols (green, blue) represent the simulation data at times $t = 3, 6$, matching perfectly with their corresponding theoretical $\mathbb{P}(x, t|x_0)$ solid lines. It is compared with the steady-state distribution (dashed line, magenta color). Panel (b): Same data plotted against the large deviation function $\phi(y)$. It clearly shows the distinct behavior of $\phi(y < v)$ and $\phi(y > v)$. The black vertical dashed lines mark the positions $x = vt$ or $y = v$. The parameters are: $x_0 = 1, v = 1, D = 0.7$.}
    \label{fig:DD_DT}
\end{figure}

From the Laplace-space propagator under TR, we can study long-time relaxation behavior. A comparison plot between the steady state $\mathbb{P}_{\text{ss}}(x)$, and the finite-time propagator $\mathbb{P}(x, t|x_0)$ is shown in Fig.~\ref{fig:DD_DT}(a) for two different times. In the limit $t \to \infty, x \to \infty$ keeping $x/t$ fixed, the propagator gives rise to a large-deviation form.
\begin{equation}
    \mathbb{P}(x, t|x_0) \sim \exp\left(-t\phi(x/t)\right),
\end{equation}
where large $x$ represents the tail of the propagator (i.e., $x > x_0$), and $\phi(y)$ is the rate function, defined as follows
\begin{equation}
    \phi(y) = -\lim_{t \to \infty} \frac{ \ln[ \mathbb{P}(y~t, t|x_0)]}{t}.
\end{equation}
Using the results derived in the appendix~\ref{app:dd-propagator}, we find that
\begin{align}\label{rateFunc}
    \phi(y= x/t) = 
    \begin{cases}
        \dfrac{(y + v)^2}{4D}, & y > v\\[0.7pc]
        \dfrac{y~v}{D}, & y < v
    \end{cases}
\end{align}
In the $(x,t)$ plane, there is thus a light cone moving with speed $v$, i.e., $x^\star(t)= v t$, and one sees that the rate function $\phi(x/t=y)$ has different behaviors in the inside and the outside of the light cone, as in Eq. \eqref{rateFunc}. While the rate function $\phi(y)$ and its first derivative are continuous at $y^\star=v$, its second derivative is discontinuous, signaling a second-order dynamical phase transition. Such a second order dynamical phase transition in the relaxation spectrum was first found in \cite{majumdar2015dynamical} for diffusion with Poissonian resetting, and also more recently, in the context of mortal Brownian motions in \cite{yerrababu2026dynamical} and switching diffusion \cite{gueneau2025large}.

\section{Conclusion}
\label{conclusion}
In this work, we investigated the existence of a steady state for a single particle undergoing arbitrary dynamics under threshold resetting (TR), motivated by threshold-driven phenomena. We showed that, whenever a steady state exists, it can be expressed as the ratio of two physically meaningful quantities: the mean local time (MLT) and the mean first-passage time (MFPT). The MFPT represents the total time taken by the particle to reach the threshold for the first time, while the MLT characterizes how this time is distributed across space. Both quantities are determined solely by the underlying dynamics. This relation provides a sufficient condition for the existence of a non-equilibrium steady state, namely the finiteness of the MFPT to the threshold.

To illustrate these results, we analyzed drift–diffusion, a Markovian process. The exact steady state under TR was obtained and expressed as the ratio of the MLT and MFPT. We also examined the temporal evolution of the moments and the MSD. In the drift-dominated regime ($\text{Pe} \gg 1$), the relaxation toward the steady state exhibits damped oscillations in both the moments and the MSD. We also studied the large-deviation $(x \sim t)$ of the propagator under TR at long times, and showed that a position undergoes a dynamical phase transition from transient to steady state.

Several directions naturally emerge from the present study. An immediate extension would be to investigate threshold resetting for a broader class of underlying stochastic processes, both heterogeneous diffusive and non-diffusive, and to generalize the analysis to $N$ random walkers \cite{bonomo2021first}, by combining the present formalism with the recent developments in \cite{biroli2026first}. The framework may also provide a useful perspective on threshold-triggered dynamics in biological systems, such as integrate-and-fire neurons \cite{burkitt2006review}. A particularly intriguing direction concerns branching dynamics: recent work has shown that when a parent cell branches into $m$ daughter cells upon threshold crossing, the population can exhibit exponential growth \cite{kumar2026branching}, whereas the $m=1$ case considered here leads to a non-equilibrium steady state. Understanding the moments of the population for general $m$ would therefore provide a natural extension of the present setting and, importantly, could be verified with experimental observations. Finally, the growing experimental interest in stochastic resetting  \cite{besga2020optimal,faisant2021optimal,tal2020experimental,altshuler2024environmental,kundu2025emulating,paramanick2024uncovering} suggests that these systems can also offer a promising platform for exploring threshold-controlled resetting dynamics. 

\section{Acknowledgments}
The numerical calculations reported in this
work were carried out on the Kamet cluster, which is
maintained and supported by the Institute of Mathematical Science’s High-Performance Computing Center. SNM acknowledges support from ANR Grant No. ANR23- CE30-0020-01 EDIPS. AP acknowledges research funding under the scheme
ANRF/ARGM/2025/001623 from ANRF, India and research support from the Department of Science and Technology, India, SERB Start-up Research Grant Number SRG/2022/000080. SNM and AP also acknowledge the International Research Project (IRP) titled ``Classical and quantum dynamics in out of equilibrium systems'' by CNRS, France. RD and AP gratefully acknowledge research support from the Department of Atomic Energy, Government of India via the Apex project ``Modelling Structure \& Dynamics of Soft Matter''.

\appendix







\section{Inverse Laplace transform of the observables in the DD case}\label{DD_appdx}
\begin{figure}[b]
    \centering
    \includegraphics[width=0.8\linewidth]{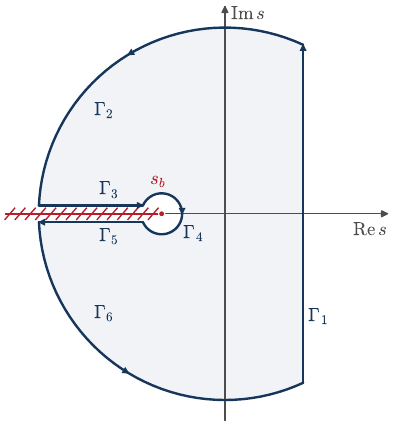}
    \caption{Contour for the Bromwich integral in the drift–diffusion process with threshold resetting (TR). The red point denotes the branch point at $s = s_b$ [see Eq.~\eqref{BP_DD}], giving rise to a branch cut along $(-\infty, s_b)$. The vertical contour $\Gamma_1$ is chosen such that all poles of the integrand (including the pole at the origin) lie to its left.}
    \label{fig:figcon1}
\end{figure}

We recall the exact expression of the mean position in Laplace space from Eq. (\ref{DD_mean_ls}) which is given by
\begin{equation}\label{mean_DD_ls_appdx}
    \langle \widetilde{x}(s)\rangle =
    \frac{1}{s}\frac{x_0}{1-\exp\!\left[\text{Pe}(1-\lambda(s))\right]}
    -
    \frac{v}{s^2},
\end{equation}
where $\lambda(s) = \sqrt{1 + {4Ds}/{v^2}}$. We can compute the exact time-domain solution analytically by using the inverse Laplace transform, which is defined via the Bromwich integral
\begin{equation}
    \langle x(t)\rangle 
    = \frac{1}{2 \pi i} \int_{\Gamma_1} ds\, e^{st}\, \langle \widetilde{x}(s)\rangle,
\end{equation}
where $\Gamma_1$ is a vertical contour in the complex $s$-plane lying to the right of all singularities (see Fig.~\ref{fig:figcon1}).
Using Cauchy's residue theorem, we deform the Bromwich contour into a closed contour consisting of segments $\Gamma_1$--$\Gamma_6$, yielding
\begin{equation}\label{cauchy}
    \frac{1}{2\pi i} \sum_{k=1}^{6} \int_{\Gamma_k} ds\, e^{st}\, \langle \widetilde{x}(s)\rangle
    = \sum \text{Res}\left[e^{st}\langle \widetilde{x}(s)\rangle\right],
\end{equation}
where the sum on the right-hand side runs over all poles enclosed by the contour.

\subsection{Singularity structure}

The square-root term in $\lambda(s)$ introduces a branch point located at
\begin{equation}\label{BP_DD}
    s_b = -\frac{v^2}{4D},
\end{equation}
with a branch cut chosen along $(-\infty, s_b]$ line.
In addition, there is a pole at $s=0$. The remaining poles arise by equating the denominator of $\langle \widetilde{x}(s) \rangle$ to zero.
\begin{equation}
    1 - \exp\!\left[\text{Pe}\left(1 - \lambda(s)\right)\right] = 0,
\end{equation}
which gives
\begin{equation}
        \sqrt{1 + \frac{4Ds}{v^2}} = 1 - \frac{4 \pi n D}{x_0 v}i,
\end{equation}
As the complex point in the RHS has a positive real part, squaring it doesn't push it out of the principal Riemann sheet.
After solving for $s$, we get the position of the poles
\begin{align}\label{eq:polesApdx}
    s_n &= -\frac{4\pi^2 n^2 D}{x_0^2} 
    - i\,\frac{2\pi n v}{x_0}, 
    \quad n \in \mathbb{Z}\\
    &= -\gamma_n - i \omega_n.
\end{align}
These poles occur in complex conjugate pairs. Their negative real part ($\gamma_n$) leads to exponential decay, while the imaginary part ($\omega_n$) generates oscillatory behavior in time.

\subsection{Residue Contributions}

The residue at $s=0$ corresponds to the steady-state contribution and reproduces Eq. \eqref{ss_mean_msd_dd}.
\begin{equation}
    \operatorname{Res}(s=0) 
    = \frac{x_0}{2} + \frac{D}{v} = \frac{x_0}{2}\left(1 + \frac{1}{\text{Pe}}\right) = \langle x\rangle_{\text{ss}},
\end{equation}
Due to $n \neq 0$ poles, the residues are
\begin{equation}
    \operatorname{Res}(s_n) 
    = -x_0 \frac{\text{Pe} - 2\pi ni}{2\pi n(\pi n + i \text{Pe)}} e^{s_n t}.
\end{equation}
Combining conjugate poles $(n,-n)$, the contribution can be written as
\begin{equation}
\operatorname{Res}(s_n) + \operatorname{Res}(s_{-n}) = \mathcal{C}_n\cos(\omega_n t+ \theta_n) e^{-\gamma_n t},
\end{equation}
where the amplitude and phase are defined as
\begin{align}
    \mathcal{C}_n &= \frac{x_0}{n \pi}\sqrt{\frac{\text{Pe}^2 + 4 n^2 \pi^2}{\text{Pe}^2 + n^2\pi^2}}, \theta_n = - \tan^{-1}\left(\frac{\text{Pe}^2 + 2n^2\pi^2}{n\pi \text{Pe}} \right).
\end{align}

\subsection{Branch Cut Contribution}

To evaluate the contribution from the branch cut (contours $\Gamma_3$ and $\Gamma_5$), we parametrize
\begin{equation}
    s - s_b = w e^{\pm i\pi}, \quad w \in (0,\infty).\nonumber
\end{equation}
Substituting further $w = D q^2/x_0^2$ helps to eliminate some square root terms in the integrand; this yields the branch-cut function appearing in Eq. \eqref{DD_mean_time},
\begin{equation}\label{H_tau}
\begin{aligned}
    \mathcal{H}(t) &= \frac{1}{2\pi i} \left( \int_{\Gamma_3} + \int_{\Gamma_5} \right)
    = -\frac{x_0 e^{-v^2t/(4D)}}{\pi}\\
    & \qquad\times\int_0^\infty dq\, \frac{q\,e^{-q^2 D t/x_0^2}}{q^2 + \text{Pe}^2}\,
    \frac{\sin q}{\cos q - \cosh \text{Pe}}.
\end{aligned}
\end{equation}
By Jordan's lemma, the integrals over $\Gamma_2$ and $\Gamma_6$ vanish. The small circular contour $\Gamma_4$ around the branch point also vanishes in the small radius limit.

\subsection{Full time-dependent result}

Collecting all contributions (pole at $s=0$, complex poles, and branch cut) and using them into Eq. \eqref{cauchy}, we obtain Eq. \eqref{DD_mean_time},
\begin{equation}\label{DD_mean_il}
\langle x(t)\rangle = \langle x\rangle_{\text{ss}}
+ 
\sum_{n=1}^\infty
\mathcal{C}_n \cos(\omega_n t+\theta_n)e^{-\gamma_n t}
- \mathcal{H}(t).
\end{equation}

\subsection{Alternate method using Laplace table}

We followed the Bromwich integral method above to show the emergence of these complex poles mathematically, which are responsible for the temporal oscillatory behavior. Otherwise, we can find $\langle x(t)\rangle$ and $\langle x^2(t)\rangle$ simply by using inverse Laplace tables; the resulting exact expressions are Eqs. \eqref{dd_il_table} and \eqref{dd_il_table2} of the main text. Although we could not find a closed-form solution, those expressions are sometimes relatively easier to compute in \textit{Mathematica} as we don't need to do numerical integration as in Eq. \eqref{DD_mean_il}.
The MSD can be obtained using this relation
\begin{equation}
\begin{aligned}
\langle\Delta x^2(t)\rangle
= \langle x^2(t) \rangle
- \langle x(t) \rangle^2
\end{aligned}
\end{equation}

\section{Exact form of drift--diffusion propagator under threshold resetting}
\label{app:dd-propagator}

We consider a drift--diffusion process on the half-line $x>0$ with constant
velocity $v>0$ towards the origin, diffusion constant $D$, and
instantaneous resetting to the starting point $x_0$ whenever the walker is
absorbed (\emph{threshold resetting}). 
The absorbing-boundary propagator in
Laplace space is given by (here, we use $x$ and the dimensionless parameter $u=x/x_0$ interchangeably for notational convenience)
\begin{equation}\label{eq:G}
\begin{split}
    \widetilde{\mathbb{G}}_{+}(x,s|x_0) &= \frac{1}{v\lambda(s)} e^{\mathrm{Pe}(1- u)} \\
    &\quad \times \Big\{ e^{-\lambda(s)\mathrm{Pe}|u - 1|} - e^{-\lambda(s)\mathrm{Pe}(u + 1)} \Big\},
\end{split}
\end{equation}
with $\lambda(s)=\sqrt{1+4Ds/v^2}$, and the corresponding first-passage time
density to the origin is
\begin{equation}
  \widetilde{\mathbb{F}}(s|x_0)
  = \exp\!\left[\text{Pe}\,\left(1-\lambda(s)\right)\right].
  \label{eq:F}
\end{equation}
The propagator under threshold resetting is as follows,
\begin{equation}
  \widetilde{\mathbb{P}}(x,s|x_0)
  = \frac{\widetilde{\mathbb{G}}_{+}(x,s|x_0)}
         {1-\widetilde{\mathbb{F}}(s|x_0)} .
  \label{eq:propApdx}
\end{equation}
The finite-time propagator is recovered by the Bromwich inversion
\begin{equation}
  \mathbb{P}(x,t|x_0)
  = \frac{1}{2\pi i}\int_{\Gamma_1} ds\; e^{st}\,
    \widetilde{\mathbb{P}}(x,s|x_0).
  \label{eq:bromwich}
\end{equation}
where $\Gamma_1$ is a vertical contour in the complex $s$-plane lying to the right of all singularities of $\mathbb{\widetilde{P}}(x, s|x_0)$.

\textbf{Note:} Throughout we restrict to $u > 1$ (i.e., $x > x_0$), so that $|u - 1| = u -1$ in
Eq.~\eqref{eq:G}. This is because we are interested in studying the large-$x$ behavior of the propagator at large times (large deviation).

\subsection{Singularity structure}

Using Eq. \eqref{eq:propApdx}, we find in the complex-$s$ plane, that $\widetilde{\mathbb{P}}(x, s|x_0)$ has (i) simple pole at $s = 0$ and other poles at the
zeros of $1-\widetilde{\mathbb{F}}$ and (ii) a branch point at
$s_{b}=-v^{2}/4D$ (due to square root present in $\lambda(s)$), with a cut along
$s\in(-\infty,s_{b}]$. Deforming $\Gamma_1$ onto the cut (see Fig.~\ref{fig:figcon1}) and collecting the
enclosed residues decomposes the propagator as
\begin{equation}
  \mathbb{P}(x,t|x_0)
  = \mathbb{P}_{\text{poles}}(x,t|x_0)
  + \mathbb{P}_{\text{BP}}(x,t|x_0),
  \label{eq:decomp}
\end{equation}
where $\mathbb{P}_{\text{poles}}$ is the sum of residues and
$\mathbb{P}_{\text{BP}}$ is the branch-cut (Hankel) integral.

\subsection{Pole contribution}

The poles solve $1-\widetilde{\mathbb{F}}(s|x_0)=0$, which gives us the same poles as in Eq. \eqref{eq:polesApdx}.
After computing the residues of these poles and combining $n$ and $-n$ terms, we get the contribution of the poles by summing over all these residues. The full expression is written below.
\begin{equation}
\boxed{
\begin{aligned}
  \mathbb{P}_{\text{poles}}(x,t|x_0)
  &= \mathbb{P}_{\text{ss}}(x)
   + \frac{2}{x_0}\,e^{-2\text{Pe}u}
     \left[e^{2\text{Pe}}-1\right] \\
  &\quad \times \sum_{n\ge 1}
     e^{-\gamma_n t}\,
     \cos\!\left(2\pi n u - \omega_n t\right).
\end{aligned}
}
\label{eq:Ppoles}
\end{equation}
where the first term (due to the $ s=0$ pole) is the non-equilibrium steady state (NESS).
\begin{equation}
  \mathbb{P}_{\text{ss}}(x)
  = \frac{1}{x_0}\bigl(1 - \exp[-2\text{Pe}]\bigr) \exp[-2\text{Pe}(u-1)],
  \quad u > 1 .
  \label{eq:Pss}
\end{equation}


\subsection{Branch-cut contribution}

\begin{figure}
    \centering
    \includegraphics[width=1\linewidth]{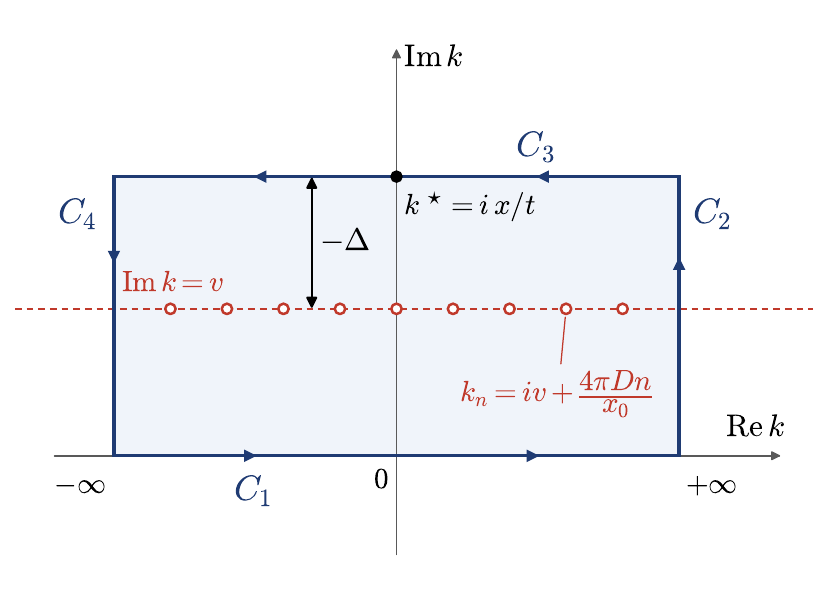}
    \caption{Positioning of the poles of $\mathcal{A}(k)$ and the saddle point $k^\star$ (from Eq. \eqref{eq:PBPfull}) on the complex k-plane. $C_1, C_2, C_3$, and $C_4$ together forms the closed contour. The poles (red markers) lie on the horizontal line $\text{Im}(k) = v$. The branch cut integral in Eq. \eqref{eq:PBPfull} is defined on contour $C_1$, whereas the saddle-path integral $I_{\text{saddle}}(t)$ is defined in the exact opposite direction to the contour $C_3$.}
    \label{fig:kPlaneContour}
\end{figure}

Parameterizing the cut by $s= s_b - k^2/(4D)$, the brach-cut integral
reduces to

\begin{equation}
\boxed{
\begin{aligned}
  &\mathbb{P}_{\text{BP}}(x,t|x_0) = \frac{1}{\pi D} \exp[\text{Pe}(1- u)] \\[0.5 pc]
  &\quad \times \int_{0}^{\infty}\! dk\, \sin(\text{Pe}\,k/v) \exp\left[-\frac{v^2 + k^2}{4D}t\right] \\[0.5 pc]
  &\quad \times \frac{\sin(\text{Pe}\,u\,k/v) - \exp({\text{Pe}})\sin(\text{Pe}(u - 1)k/v)}
               {1 - 2\exp({\text{Pe}})\cos(\text{Pe}\,k/v) + \exp({2\text{Pe}})},
\end{aligned}
}
\label{eq:PBPhalf}
\end{equation}
where the denominator is $\bigl|1-e^{(1+ik/v)\text{Pe}}\bigr|^{2}$. As the integrand is an even function, we can extend the range to $(-\infty,\infty)$ by adding one extra $1/2$ factor, and writing sines and cosines in terms of the exponential yields the compact
form
\begin{equation}
  \mathbb{P}_{\text{BP}}(x,t|x_0)
  = \mathcal{C}\int_{-\infty}^{\infty} dk\;
    \mathcal{A}(k)\,e^{S(k)},
  \label{eq:PBPfull}
\end{equation}
with

\begin{align}
  \mathcal{A}(k) &= \frac{\sin\!\left(\text{Pe}~k/v\right)}
                         {1-\exp\!\left[\text{Pe}(1 + i k/v)\right]}, \\
  \mathcal{C} &= \frac{1}{2\pi i D}\,
                 \exp\!\left[\text{Pe}(1 - u) - v^2 t/(4D)\right], \\
  S(k) &= -k^2 t/(4 D) + i~\text{Pe}~ u ~(k/v).
\end{align}
The phase $S(k)$ is exactly Gaussian, so its stationary point is the only
saddle: $\partial_k S=0$ gives $k^{\star}=i x/t$. The factor $\mathcal{A}(k)$
contributes simple poles where $\exp[\tfrac{x_0}{2D}(v+ik)]=1$, i.e.\ at
$k_{n}=iv+\tfrac{4\pi D n}{x_0}$. These poles sit at height $\operatorname{Im}k=v$,
while the steepest-descent contour through the saddle is the horizontal line
$\operatorname{Im}k=x/t$. Deforming the real-$k$ axis up to this line therefore
sweeps across the poles only when $x/t>v$ (shown in Fig\ref{fig:kPlaneContour}); by the residue theorem,
\begin{align}
  &\mathbb{P}_{\text{BP}}(x,t|x_0)-I_{\text{saddle}}(t)
  =\nonumber\\
  &\qquad\begin{cases}
    \displaystyle 2\pi i\sum_{n=-\infty}^{\infty}\operatorname{Res}_{k = k_n}\left[\mathcal{C}\mathcal{A}(k)e^{S(k)} \right],
      & x/t>v,\\[4mm]
    0, & x/t<v,
  \end{cases}
  \label{eq:bpcase}
\end{align}
where we have used the fact that contribution due to $\int_{C_2}$, and $\int_{C_4}$ vanishes as $|\text{Re}(k)| \to \infty$ in both of the integration. $I_{\text{saddle}}(t)$ denotes the same integrand as $\mathbb{P}_{\text{BP}}$ evaluated along the saddle
path. 
A short calculation shows that the residue series
reconstructs the pole contribution (Eq. \eqref{eq:Ppoles}) exactly with a $-$ve sign:
\begin{equation}
  2\pi i\sum_{n=-\infty}^{\infty}\operatorname{Res}_{k = k_n}\left[\mathcal{C}\mathcal{A}(k)e^{S(k)}\right]
  = -\,\mathbb{P}_{\text{poles}}(x,t|x_0).
  \label{eq:rescancel}
\end{equation}
Combining Eqs.~\eqref{eq:decomp}, \eqref{eq:bpcase} and \eqref{eq:rescancel},
the pole term cancels whenever $x/t>v$, so that
\begin{equation}
  \mathbb{P}(x,t|x_0)
  = \Theta\!\left(v-\frac{x}{t}\right)\mathbb{P}_{\text{poles}}(x,t|x_0)
  + I_{\text{saddle}}(t).
  \label{eq:final_prop}
\end{equation}
Note: For $x > v~t$, the steady-state contribution due to the pole at the origin vanishes, so the steady state is not yet achieved at this time $t$ for this spatial region, and the relaxation front is moving with a velocity $v$.

\subsection{Evaluation of the saddle-path integral}

Shifting $k=i x/t+\rho$ and using that $S$ is quadratic gives the exact result
$S(i x/t+\rho)=-x^{2}/4Dt-\rho^2t/4D$, hence
\begin{equation}\label{eq:saddleInt}
\begin{split}
  I_{\text{saddle}}(t)
  &= \mathcal{C}\,e^{-x^{2}/4Dt} \\
  &\quad \times \int_{-\infty}^{\infty} d\rho\;
    \mathcal{A}(\rho+i x/t)\,e^{-\rho^2 ~t/4D}.
\end{split}
\end{equation}
where position of the poles due to $\mathcal{A}(\rho + ix/t)$ with respect to $\rho$ is now as follows
\begin{equation}
    \rho_n = k_n - i\frac{x}{t} = i\left(v - \frac{x}{t}\right) + \frac{4\pi Dn}{x_0}.
\end{equation}
We can compute the integral in Eq. \eqref{eq:saddleInt} in the long-time limit using the saddle point method\cite{wong2001asymptotic}, but we have to be careful when a pole $\rho_p$ in the integrand overlaps with the saddle point $\rho = 0$. The pole of $\mathcal{A}$ nearest the saddle is the $\rho_{n = 0}$, located at
$\rho_{p}= \rho_{n = 0}=i\Delta$ with
\begin{equation}
  \Delta \equiv v - \frac{x}{t}=\frac{vt - x}{t}.
\end{equation}
When $\Delta\to0$, i.e., $x \to vt$, this pole pinches the contour and overlaps with the saddle point, so we isolate it by writing
\begin{equation}
\begin{aligned}
  \mathcal{A}(\rho+i x/t)
  &= \frac{\operatorname{Res}(\rho_p)}{\rho-\rho_p}
  + \mathcal{A}_{\text{reg}}(\rho), \\
    \operatorname{Res}(\rho_p)
  &= -\frac{2D}{x_0}\sinh\!\left(\text{Pe}\right),
\end{aligned}
\end{equation}
which splits the integration in Eq. \eqref{eq:saddleInt} into two parts, one containing the singular part of $\mathcal{A}$, and another one containing the regular part. We are going to compute each of them next.
\begin{equation}\label{eq:saddleSplit}
    I_{\text{saddle}}(t) = I_{\text{sing}}(t) + I_{\text{reg}}(t)
\end{equation}

\noindent\textbf{a. Singular part}

With the standard Gaussian--Cauchy integral (real $b$), we know that\cite{abramowitz2006handbook}
\begin{equation}
\begin{split}
  \int_{-\infty}^{\infty} d\rho\;
  \frac{e^{-\rho^2t/4D}}{\rho-i b}
  &= i\pi\,\operatorname{sgn}(b)\,
    e^{b^{2}t/4D} \operatorname{erfc}\!\left(|b|\sqrt{\tfrac{t}{4D}}\right).
\end{split}
\end{equation}
Applied with $\rho_p=i\Delta$, i.e.\ $b=\Delta$ (so that
$\operatorname{sgn}(b)=\operatorname{sgn}(vt-x)$, and 
$|b|\sqrt{t/4D}=|vt-x|/\sqrt{4Dt}$)
one obtains the exact contribution of the singular part
\begin{equation}
\boxed{
\begin{aligned}
  I_{\text{sing}}(t)
  &= -\frac{1}{x_0}\sinh\!\left(\text{Pe}\right)
     \exp\!\left[\text{Pe}(1-2u)\right] \\
  &\quad \times
     \operatorname{erfc}\!\left(
       \frac{|x-vt|}{\sqrt{4Dt}}
     \right)
     \operatorname{sgn}(vt-x).
\end{aligned}
}
\label{eq:Ising}
\end{equation}
\noindent\textbf{b. Regular part}

Since $\mathcal{A}_{\text{reg}}$ is analytic at $\rho=0$, the long-time limit
follows from Laplace's method\cite{bender1999advanced},
$\int d\rho\,\mathcal{A}_{\text{reg}}(\rho)e^{-\rho^2 t/4D}\simeq \mathcal{A}_{\text{reg}}(0)\int e^{-\rho^2 t/4D} d\rho =  
\mathcal{A}_{\text{reg}}(0)\sqrt{4\pi D/t}$, with
\begin{equation}
\begin{split}
  \mathcal{A}_{\text{reg}}(0)
  &= \mathcal{A}(i x/t)+\frac{\operatorname{Res}(\rho_p)}{i\Delta} \\
  &=i\left[\frac{\sinh(xx_0/2Dt)}{1-e^{\text{Pe}\Delta/v}}+\frac{v\sinh(\text{Pe})}{\text{Pe}\,\Delta}\right].
\end{split}
\end{equation}
Collecting the prefactors (present in Eq. \eqref{eq:saddleInt}) gives

\begin{equation}
\boxed{
\begin{aligned}
  I_{\text{reg}}(t)
  &= \frac{1}{\sqrt{\pi D t}}\,
     \exp\!\left[
       \text{Pe} - \frac{(x+vt)^2}{4Dt}
     \right] \\
  &\quad \times
     \left[
       \frac{\sinh\!\left(xx_0/2Dt\right)}
            {1-e^{\text{Pe}\Delta/v}}
       + \frac{v\sinh\!\left(\text{Pe}\right)}
              {\text{Pe}\,\Delta}
     \right].
\end{aligned}
}
\label{eq:Ireg}
\end{equation}
Eqs.~\eqref{eq:final_prop}, \eqref{eq:saddleSplit}, \eqref{eq:Ising} and \eqref{eq:Ireg} constitute the
full propagator
\begin{equation}\label{asympForm}
    \mathbb{P}(x,t|x_0)
  = \Theta\!\left(v-\frac{x}{t}\right)\mathbb{P}_{\text{poles}}
  + I_{\text{reg}}(t) + I_{\text{sing}}(t).
\end{equation}

Note: To evaluate $I_{\text{reg}}(t)$ in Eq. \eqref{eq:Ireg}, we have taken the long-time limit. If we keep it in integral form without doing any asymptotic analysis, the above equation \eqref{asympForm} is valid for any finite time $t$.

\subsection{Large Deviation Form}

After studying the large-time asymptotics of the different terms present in $\mathbb{P}(x, t|x_0)$ [Eq. \eqref{eq:final_prop}], we can find the large deviation form of the propagator in the limit $t \to \infty, x \to \infty$ keeping $x/t$ fixed.
\begin{equation}
    \mathbb{P}(x, t|x_0) \sim \exp\left(-t\phi(x/t)\right),
\end{equation}
where $\phi(y)$ is the rate function, defined as follows
\begin{equation}
    \phi(y) = -\lim_{t \to \infty} \frac{ \ln[ \mathbb{P}(y~t, t|x_0)]}{t}.
\end{equation}
Using the results derived in the previous sections (see Eq. \eqref{asympForm} and corresponding expressions of each term), we find that
\begin{align}
    \phi(y= x/t) = 
    \begin{cases}
        \dfrac{(y + v)^2}{4D}, & y > v\\[0.7pc]
        \dfrac{y~v}{D}, & y < v
    \end{cases}
\end{align}
The second derivative $\phi^{\prime\prime}(y)$ is discontinuous at $y = v$. This indicates that a position $x$ goes through a second-order dynamical phase transition from a transient state to a steady state at a characteristic time $ t = x/v$.

\bibliographystyle{unsrt}
\bibliography{references}

\end{document}